\documentclass[aps,rmp,preprint,groupedaddress,showkeys]{revtex4-2}

\usepackage{graphicx}
\usepackage{bm}
\usepackage{array}
\usepackage{amsmath}
\usepackage{booktabs}
\usepackage{multirow}
\usepackage{makecell}
\usepackage{geometry}
\usepackage{enumitem}
\usepackage{changepage}
\usepackage{hyperref}

\begin{document}


\title{From Novice to Expert in Cloud Physics: a Network-Based Analysis of Learner Understanding}


\author{}
\affiliation{}

\author{Julien-Pooya Weihs}
\email{Corresponding author: julien-pooya.weihs@uib.no}
\affiliation{Geophysical Institute, University of Bergen, Bergen, Norway}
\affiliation{Bjerknes Centre for Climate Research, Bergen, Norway}

\author{Vegard Gjerde}
\affiliation{Department of Physics and Technology, University of Bergen, Bergen, Norway}

\author{Helge Drange}
\affiliation{Geophysical Institute, University of Bergen, Bergen, Norway}
\affiliation{Bjerknes Centre for Climate Research, Bergen, Norway}


\date{\today}

\begin{abstract}
Understanding how learners conceptualize complex scientific systems remains a key challenge in geoscience education. We investigate the evolution of conceptual understanding in cloud physics among 153 learners, ranging from bachelor students to disciplinary experts and representing diverse academic backgrounds across STEM. To do so, we trace how knowledge structures differ across levels of experience using metrics from a cross-sectional network analysis. The analysis characterizes the quantitative and qualitative dimensions of the epistemological shift that learners experience as they mature in their understanding of the discipline. We show that in their description of the life-cycle of a cloud, they progressively transition from the general physics of the water cycle to detailed descriptions of cloud microphysical processes.
A triangulation of data sources with a panel of experts complements and confirms the analysis. The results can assist lecturers in structuring their teaching toward higher levels of understanding and enable students to anticipate the key complexities and conceptual challenges in the field during their learning process. Furthermore, the generic nature of the analysis can be transferred to a wide range of disciplines.
\end{abstract}

\keywords{Atmospheric physics; Cloud physics; Conceptual understanding; Knowledge structures; Graph network analysis}

\maketitle

\section{Introduction\label{sec:intro}}

Clouds significantly influence the Earth's water cycle and climate, making their understanding critical to scientific and societal discussions on climate change \cite{Forster2021, Stephens2019}. Cloud physics investigates the scientific processes governing the life-cycle of clouds, from their formation to dissipation, addressing key topics such as the effect of aerosols on droplet nucleation \cite{Martinsson1999, Moehler2007}, conditional droplet growth mechanisms \cite{Li2019}, and particle--turbulence interactions \cite{Vaillancourt2000}. The field integrates advances in atmospheric dynamics, physical meteorology, chemistry, and thermodynamics \cite{Ambaum2020, Lohmann2016, Wallace2006}, and remains central to addressing persistent uncertainties in climate modeling \cite{Hansen2023, Morrison2020, Pathak2020}. Cloud physics education is therefore a key component of atmospheric physics education and, more broadly, geoscience education \cite{Cervato2018, Forster2021, Stephens2019}.

Research in cloud physics education has predominantly focused on learning difficulties related to singular concepts, such as vapor pressure, evaporation, condensation, and droplet nucleation processes \cite{Gopal2004, Haaland2010, Rappaport2009, Weihs2025}. However, there is limited investigation into how these individual concepts are interconnected within the learners' knowledge structures. Analyzing relationships between concepts could offer deeper insights into how understanding develops with more experience in the field. Existing approaches in atmospheric physics education have not yet explicitly utilized methods designed to represent and analyze these interconnected conceptual relationships. In addition, studies have shown that expertise relies on both the possession of extensive content knowledge, and the appropriate judgment on how to organize it \cite{FarringtonDarby2006}. Building on this, Goldwater and Schalk \cite{Goldwater2016} have argued that scientific knowledge is inherently relational, rather than comprised of isolated features. This motivates us to consider the analytical framework of graph theory, centered around conceptual objects and the relationships between them, to explore learner understanding --- and its development --- in cloud physics.

Graph theory is an established field of mathematics, but has been sparsely applied to education research \cite{Giabbanelli2023, Podschuweit2020, Selinski2014, Turkkila2022}. However, it has shown promise in quantifying the differences between novices and experts by characterizing and assessing the quality of written explanations \cite{Wagner2022, Wagner2020, Wagner2023}, comparing graphical representations of the discipline \cite{Koponen2008, Koponen2010}, or analyzing interactions during laboratory sessions \cite{Kontro2020} or classrooms \cite{Bruun2013, Commeford2021}.

Previous educational studies have rarely incorporated graph--edge directionality or weights, limiting their ability to represent conceptual relationships for groups of learners comprehensively. We use weighted directed graphs to explore this gap, and operationalize them as \textit{conceptual networks}: nodes represent scientific concepts, and directed edges represent causal relationships between concepts according to the participants, building on the works of Koponen and Pehkonen \cite{Koponen2010}, Koponen and Nousiainen \cite{Koponen2014}, Siew \cite{Siew2020}, and Wagner \cite{Wagner2022}; the edges' weights correspond to the number of times they are expressed by a group of participants. This methodological extension enables the analysis of conceptual networks at both graph- and node-levels, providing a finer-grained picture of how understanding develops towards expertise.

Based on the above, the following research questions are addressed:
\begin{enumerate}[itemsep=0pt,parsep=0pt,topsep=0pt]
    \item How does understanding in cloud physics vary across levels of disciplinary experience, as derived from weighted directed graphs?
    \item Which distinct patterns can be identified in learners' knowledge structures by using graph network metrics, and how can these patterns inform educational practices?
\end{enumerate}

To answer the first question, we introduce a graph-based method for tracing conceptual understanding, and describe its development with increased experience. The second question is explored by analyzing and comparing the conceptual networks of different groups of learners in cloud physics, deriving possible implications for teaching. Preliminary results of this study have been presented at international conferences \cite{Weihs2024ardist, Weihs2024ecer}.

\subsection{Theoretical framework\label{subsec:theoreticalframework}}

\subsubsection{Conceptual understanding and knowledge structures\label{subsubsec:understandingandknowledge}}

The analysis of conceptual understanding has emerged in the learning sciences as a key line of inquiry for improving teaching and learning practices. It focuses on the complexities and difficulties learners encounter in constructing meaning, in contrast with less problematic tasks such as the acquisition of skills and facts \cite{diSessa2014}. Within this tradition, research on conceptual understanding has addressed fundamental questions about the nature of knowledge \cite{Pines1986}, the conditions and processes through which it is acquired \cite{Vosniadou1992, Wickman2012}, and strategies that promote deep understanding \cite{Carver2005}. In this study, understanding is viewed through the epistemological accounts of Kvanvig \cite{Kvanvig2003} and Elgin \cite{Elgin2012}. Kvanvig distinguishes understanding from the accumulation of propositional knowledge: understanding concerns a body of information and requires grasping the explanatory, logical, or causal relationships among its elements. Elgin similarly characterizes understanding as an `epistemic achievement' arising when cognitive elements form a coherent structure that supports explanation and inference. Importantly, both accounts allow understanding to coexist with some inaccurate or incomplete elements, provided that the broader structure remains sufficiently coherent and grounded. This distinction is particularly relevant for studying developing learners, whose conceptual structures may contain scientifically incomplete or incorrect elements while nevertheless exhibiting meaningful forms of organization.

This structural view also resonates with cognitive perspectives on expertise and learning, where advanced understanding is characterized by more domain-specific knowledge organized hierarchically around central principles \cite{FarringtonDarby2006}. Similarly, scientific concepts are often understood through their relationships to other concepts, making relational structure an important component of successful science learning \cite{Goldwater2016}. These perspectives collectively suggest that learning involves the reorganization of conceptual relationships rather than a mere accumulation of new facts, a view shared by several models of conceptual change research \cite{Duit2003, diSessa2018, Vosniadou2014}.

Our framework is situated within a radical constructivist view of learning \cite{Kalina2009, VonGlasersfeld2013}, according to which learners actively construct knowledge through processes of personal meaning-making. Together, these perspectives motivate our treatment of learners’ conceptual understanding as a structured system of interconnected elements whose organization may evolve with disciplinary experience. A more detailed discussion of the epistemological and cognitive foundations of this framework is provided by Weihs \cite{Weihs2026phd}.


\subsubsection{Prior investigations of understanding and knowledge\label{subsubsec:priorinvestigations}}

Across science education, numerous studies have examined how learners’ conceptual understanding forms and evolves, with particular attention to persistent misconceptions, knowledge integration, and instructional approaches to promote conceptual development \cite{Bao2021, Cervato2018, Francek2013, Herbert2006, Mills2016, Santini2018, Schoon1992}. In atmospheric physics education research, such efforts have led to the creation of assessment tools like the `Fundamentals in Meteorology [concept] Inventory' to gauge conceptual understanding in meteorology \cite{Davenport2019FMI, Davenport2015}, and to the development of new instructional approaches aimed at fostering deeper understanding \cite{Davenport2019, Persson2010, Marshall2025}.

To investigate learners’ knowledge structures, research on conceptual understanding has employed a range of representations, including concept maps \cite{Cimolino2004, Cutrer2011, Hasemann1995, Jonassen2005, Kapuza2020, Koponen2008, Novak2006, Novak2008, Watson2016, Weihs2024ardist}, system modelling tools \cite{Jonassen2005}, conceptual frameworks \cite{Dai2019}, element maps \cite{Wagner2020, Wagner2023}, causal maps \cite{Giabbanelli2023}, and cognitive maps \cite{Oezesmi2004}.
All these representations share a structural resemblance to graph networks, with elements interconnected via relationships. Despite this commonality, the use of network analysis in science education has generally been motivated by analytical considerations rather than by an explicit theoretical discussion on why conceptual understanding should itself be represented as a network. In the following subsection, we address this gap by proposing a theoretical framework that maps network representations to epistemological and cognitive theories of understanding, thereby providing a conceptual foundation for the adopted methodology.


\subsubsection{Network-based framework for understanding and knowledge\label{subsubsec:networkframework}}

Building on the theoretical perspectives outlined above, we propose a network-based framework for conceptual understanding and knowledge that integrates the epistemological accounts of Kvanvig \cite{Kvanvig2003} and Elgin \cite{Elgin2012} with cognitive perspectives on learning and expertise \cite{FarringtonDarby2006, Goldwater2016}. The proposed correspondence is as follows: \textit{concepts} (such as scientific entities, processes, or mechanisms) are represented as nodes; edges encode the \textit{knowledge} about the relationships (causal, sequential, associative, etc.) between them; the resulting complete network constitutes an individual's \textit{knowledge structure} (or \textit{conceptual structure}) for a given topic; and \textit{conceptual understanding} is interpreted as a property emerging from coherent and densely connected subgraphs of the network. Hence, conceptual change is modelled as the reorganization of the regions of highest density and connectivity within a knowledge structure, and \textit{expertise} is quantified by the properties of expert-constructed networks.
Our framework is further informed by recent work suggesting that conceptual knowledge can be `reasonably' represented as a network, and that increasing expertise is associated with greater integration and interconnectedness of knowledge structures \cite{Siew2020, Podschuweit2020, Gupta2010}.


\subsection{Methodological framework\label{subsec:methodologicalframework}}

\begin{figure}
\includegraphics[scale=0.48]{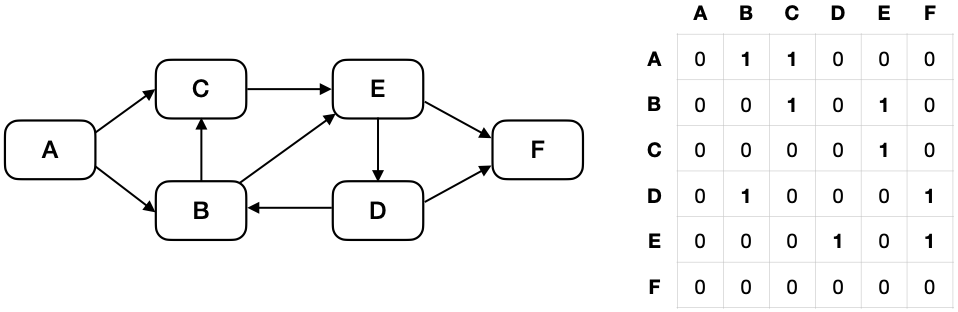}
\caption{Directed graph (left) and its associated adjacency matrix (right), inspired by Soularidis \cite{Soularidis2021}. An arrow from a node `X' to a node `Y' is encoded by a non-zero value in the cell on the $X^{th}$ row and $Y^{th}$ column of the matrix.\label{fig:figure1}}
\end{figure}

A graph is a structure consisting of a set of objects (\textit{nodes}), where some pairs are related via connections (\textit{edges}). When the edges have a direction, the graph is called \textit{directed}. When the edges are attributed a numerical value, the graph is called \textit{weighted}. Mathematically, a graph carries the same information as an adjacency matrix, where each entry represents a connection from an element of a row to an element of a column: $0$ denotes no connection, whereas $n>0$ represents a connection with weight $n$. Figure \ref{fig:figure1} illustrates a simple directed graph with maximum edge weight $1$, and its associated adjacency matrix.

Graphs have proven useful in characterizing systems, from biological systems to social networks, and various techniques and metrics have been developed to understand and analyze them \cite{Newman2003}. In educational research, graph structures are analyzed using two categories of metrics: graph-level metrics, which describe the overall structure and properties of the entire graph, and node-level metrics, which assess individual nodes and their role within the graph \cite{Giabbanelli2023, Podschuweit2020, Siew2020, Thurn2020, Wagner2020, Wang2022}. Below, we review some of the metrics used in our study and their meaning in this working context, with further details and examples provided in Supplementary File 1, available to the reader upon request.

\subsubsection{Graph-level metrics\label{subsubsec:graphlevelmetrics}}

\textit{Density} is the ratio between the existing edges and the theoretical maximum number of edges that could exist between all possible nodes of a graph. The \textit{shortest path} is computed by finding the directed path between two nodes such that the sum of the weight of its constituent edges is minimized. The \textit{diameter} is the longest of all shortest paths connecting any two nodes. The \textit{intertwinement}, or diameter-to-nodes ratio, quantifies the level of compactness of a graph, and has been well described by Wagner et al. \cite{Wagner2020}; a low value translates into a high complexity of the graph.

\subsubsection{Node-level metrics\label{subsubsec:nodelevelmetrics}}

Specific questions about a graph can be answered by computing centrality measures on its nodes. The \textit{betweenness centrality} measures how often a node lies on the shortest paths between all other nodes. It thus identifies nodes that act as bridges between other nodes in the network \cite{Thurn2020, Wagner2020}. 
\textit{Degree centrality} measures the number of connections a node has to other nodes in the network, indicating how many neighbors a node possesses \cite{Wang2022}. Nodes with high degree centrality are amongst the most connected within the network and represent concepts that have the most predecessors and successors. A further level of distinction can be made between the \textit{in-degree}, measuring the number of incoming edges, and the \textit{out-degree}, measuring the number of outgoing edges for a node. Viewed through our theoretical lens, betweenness and degree centrality capture aspects of coherence and density, respectively, and are interpreted as indicators of concepts around which conceptual understanding develops.

The \textit{eigenvector centrality} measures the influence of a node based on the influence of its neighbors: it is called \textit{left} when based on the predecessors and \textit{right} when based on the successors of the node. It assigns relative scores --- determined by the dominant eigenvector of the network's adjacency matrix \cite{Ruhnau2000} --- to all nodes in the network based on the idea that a node is important if it is connected by or to other important nodes \cite{Frost2024}. In our directed networks, nodes with high eigenvector centrality are therefore particularly important towards the end (left) or the beginning (right) of the life-cycle of a cloud. Within our theoretical framework, eigenvector centrality is interpreted as a measure of a concept's contextualized importance within the learner's entire representation of the discipline.

\subsubsection{New metrics: agreement score and timeline score\label{subsubsec:newmetrics}}

Expanding on degree centrality, we introduce in this study a new node-level metric, the \textit{agreement score}, which captures the clarity of consensus around a node’s connections in a weighted graph. Defined as the ratio of total edge weight to degree (i.e., people-per-edge), a high agreement score indicates strong consensus on a node’s connections, whereas a low score suggests greater variability in how participants link to or from that concept. We calculate separate \textit{agreement-in} and \textit{agreement-out} scores based on in-degree and out-degree, respectively, and assigning a default value of $1.0$ when either is zero. The ratio of agreement-out to agreement-in serves as a \textit{timeline score}, reflecting whether consensus is stronger on a node’s role as a source (agreement-in) or a target (agreement-out).

\subsubsection{Statistical significance\label{subsubsec:statisticalsignificance}}

This study compares computed metrics between learner groups, raising the need to assess statistical significance.. We use a permutation-based approach ($1000$ iterations) to generate empirical null distributions \cite{Edgington2007}, from which we calculate $z$-scores and $p$-values. This allows us to evaluate whether observed differences exceed what would be expected by chance. We report results in the format (metric $|$ $z$-score $|$ $p$-value), using standard thresholds for significance (e.g., $|z|\geq 1.96$ and $p<0.05$ for $95\%$ confidence).

\section{Methods\label{sec:methods}}

\subsection{Setting and population\label{subsec:settingpopulation}}

We collected hand-drawn concept maps from $153$ participants across five Norwegian universities (Nov 2022-Sep 2023) and one Swiss university (Jul 2024), spanning bachelor to expert level. Norwegian institutions were selected for convenience and disciplinary relevance, whereas the Swiss institution was included as an internationally recognized research group in cloud physics, allowing us to examine whether patterns in conceptual structures were consistent across distinct academic environments. Participants were recruited during lectures or academic events and given 10-15 minutes to complete the task.

The participants were instructed as follows: `Describe the life-cycle of a cloud, from the early conditions for formation to their dissipation'. We asked them to draw and label nodes representing scientific concepts and connect them with unlabeled directed arrows wherever they deemed relevant. 
While we call the resulting figures `concept maps', we note that our use of the term slightly differs from that of Novak and Cañas \cite{Novak2006}, who employed labeled edges. In this study, the maps are more akin to learner-generated conceptual networks, as studied amongst others by Siew \cite{Siew2020} and Podschuweit and Bernholt \cite{Podschuweit2020}.

Participants self-reported their disciplinary field, academic level, and experience with cloud physics. For the latter, they provided 1-2 sentences describing their academic exposure to the topic. The resulting texts were read by the authors, who inductively identified four groups based on increasing exposure to the discipline: \textit{Novice} (no academic exposure to the discipline), \textit{Adept} (introductory university experience), \textit{Proficient} (dedicated and more specific studies of the field), and \textit{Expert} (intensive activity in the discipline). The participants represented multiple STEM disciplines, including \textit{Geology} ($29\%$), \textit{Meteorology} ($29\%$), \textit{Physics} ($13\%$), and \textit{Geophysics} (including oceanography, air-sea interactions, etc.) ($12\%$), providing a broader view on how cloud physics is conceptualized across fields that intersect with atmospheric sciences.

\begin{figure}
\includegraphics[scale=0.36]{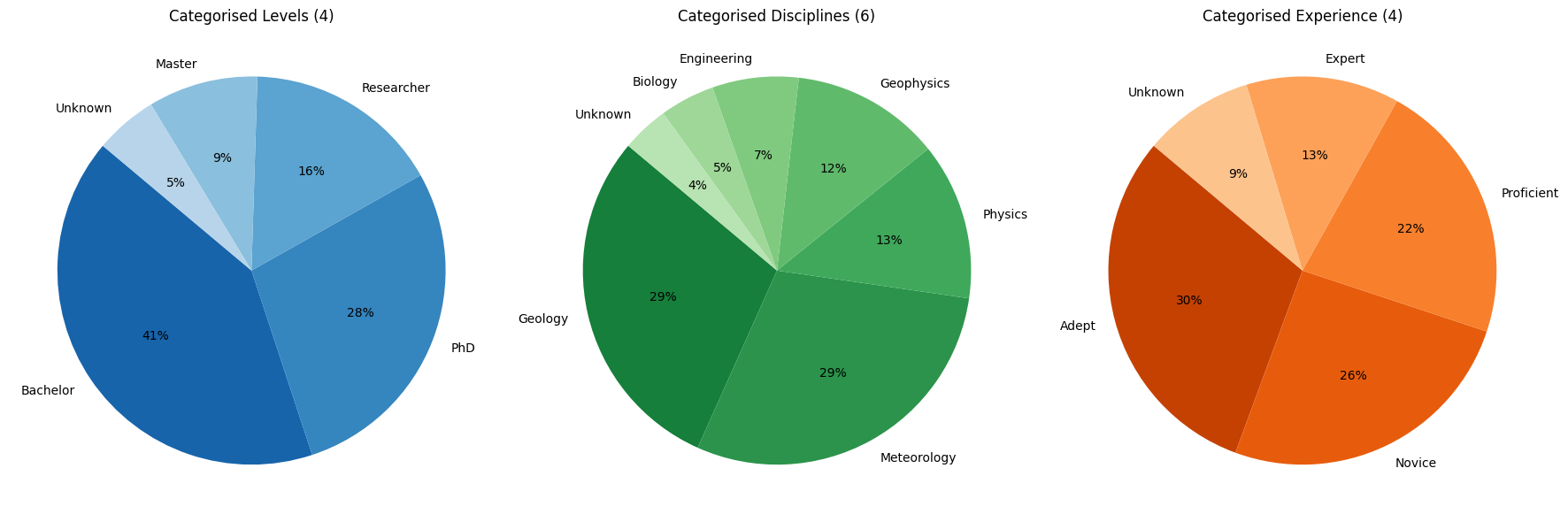}
\caption{Description of the participants according to their academic level (left, in blue), disciplinary field (middle, in green), and experience (right, in orange), inspired by Weihs and Drange \cite{Weihs2024ardist}.\label{fig:figure2}}
\end{figure}

A statistical description of the sample population is displayed in Figure \ref{fig:figure2}. The full set of self-assessments and groupings by experience, as well as a full list of included disciplines and academic levels is provided in Supplementary File 2, available to the reader upon request.

\subsection{Data pre-processing\label{subsec:datapreprocessing}}

Given the richness of the data, standardized procedures were implemented for cleaning and processing. While we aimed to remain as close as possible to the collected data, some of these choices could affect the validity and reliability of the findings. At all stages of the data cleaning and pre-processing, communication between our group members ensured that no choices were made in isolation and reflect a form of triangulation across researchers’ perspectives.

We translated all responses in Norwegian into English, whenever relevant, and decoded the handwritings. We then processed varied inputs from the participants (keywords, sentences, pictograms) into concepts. If an edge was labeled despite the instructions, those labels were either ignored (when confirming a logical or causal link, for instance `influences' or `generates'), or converted into a concept (when representing one, for instance `cools adiabatically' becoming `Adiabatic Cooling') and inserted between the two nodes linked by the initial labeled edge.

\begin{figure}
\includegraphics[scale=0.5]{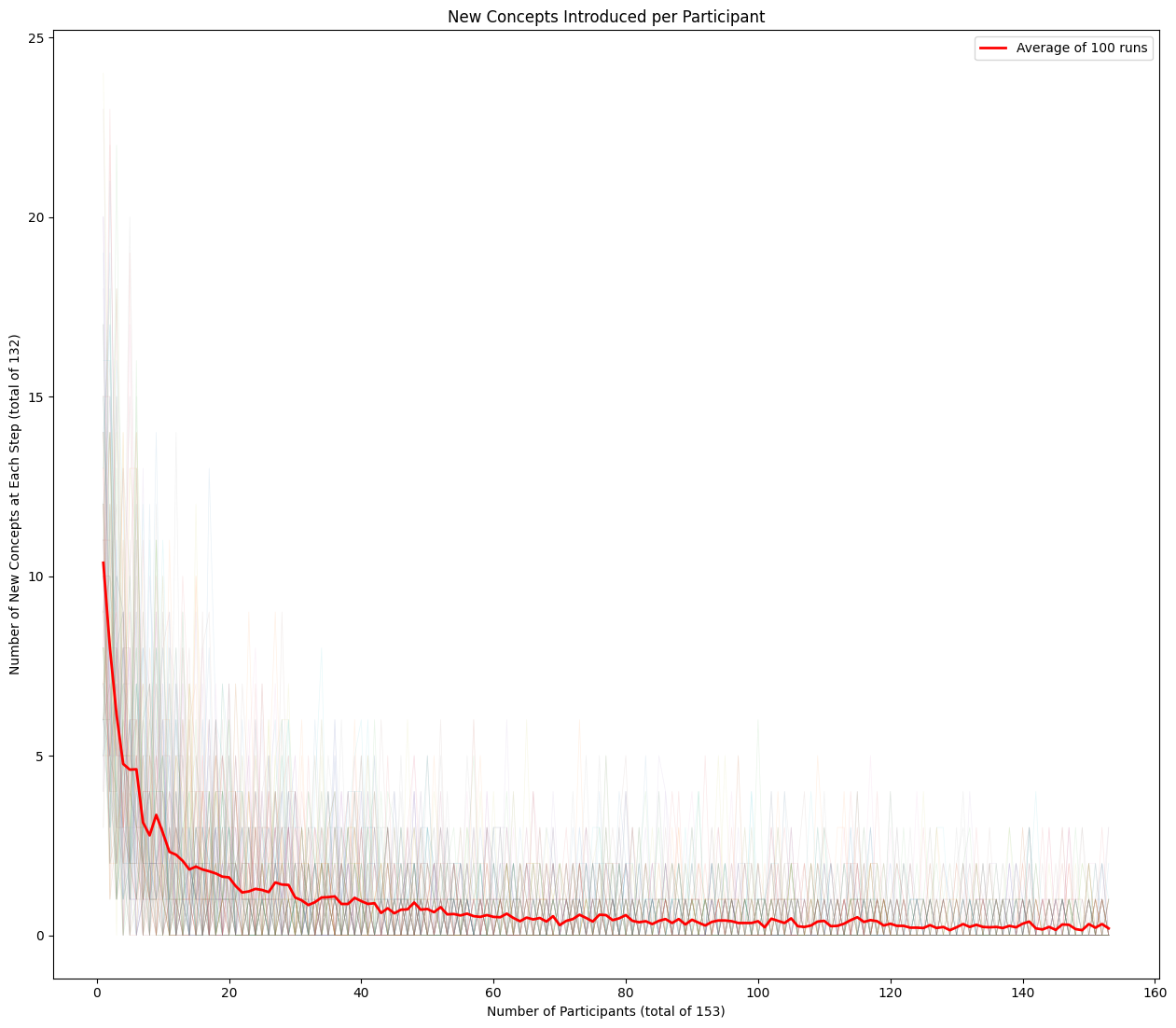}
\caption{Saturation plot of the number of new concepts introduced at each step as additional participant graphs are sequentially incorporated. For each of $100$ random orderings of the participants, the number of previously unseen concepts introduced by each new participant is tracked. Thin lines show individual runs, whereas the thick red line represents the average across all runs. The decreasing trend indicates diminishing returns in conceptual diversity as more participants are included.\label{fig:figure3}}
\end{figure}

The pre-processed data was then coded according to thematic content analysis, using an inductive approach to capture latent meaning \cite{Clarke2017, Feig2011}. An iterative dictionary of codes, available in Supplementary File 2, was created. Multiple reanalyses of the data were conducted until the codes were stable. Whenever codes showed insufficient internal coherence or were not clearly distinguishable from others, they were refined, merged, or divided, to improve clarity and consistency. The results were then reviewed and discussed among the authors. The pre-processed data as well as the coded data are presented in Supplementary File 3, available to the reader upon request, together with the the participants’ adjacency matrices associated with the coded data.
The thematic analysis approached saturation \cite{Oezesmi2004}, with only a few new codes appearing as additional data was analyzed (Figure \ref{fig:figure3}). This was verified across $100$ random orderings of maps, with the number of new concepts introduced at each step averaged and plotted.

\begin{figure}
\includegraphics[scale=0.3]{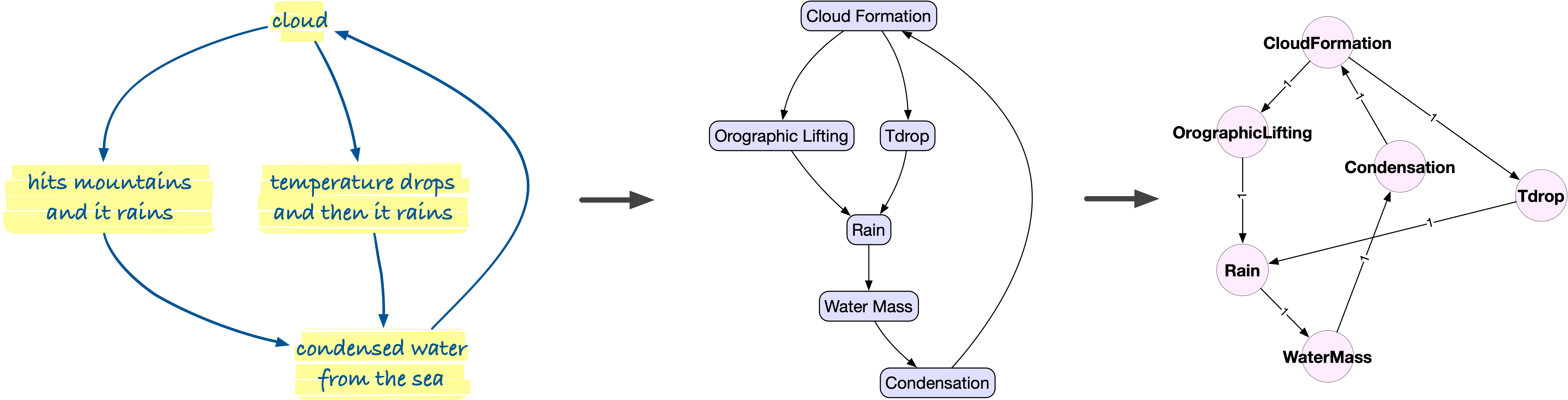}
\caption{Example of preliminary data processing progression using thematic analysis: hand-drawn data (left), coded concept map (middle), graph structure associated with adjacency matrix (right). The labels on the edges of the graph structure represent their weights (here, all of value $1$).\label{fig:figure4}}
\end{figure}

Each concept map was then converted into a graph by constructing an adjacency matrix in Python. The concepts of each participant’s map were transformed into nodes, and the arrows between them were transformed into directed edges of weight $1$. The complete conversion process from concept maps to graph structures is illustrated in Figure \ref{fig:figure4}.

\subsection{Research design\label{subsec:researchdesign}}

Graph-level metrics were calculated for each participant, following a method outlined by Wagner et al. \cite{Wagner2020} and Wagner and Priemer \cite{Wagner2023}, and tested in our preliminary study \cite{Weihs2024ardist}. We then created boxplots of the metrics according to the participants’ disciplinary field, academic level, and cloud physics experience. Categorizing the participants based on their level of experience in cloud physics revealed the largest variability in graph metrics, motivating grouping the data into four groups: \textit{Novice}, \textit{Adept}, \textit{Proficient} and \textit{Expert}, see Figure \ref{fig:figure5}.

We also combined the \textit{Novice} and \textit{Adept} groups into a broader \textit{Beginner} group, and similarly the \textit{Proficient} and \textit{Expert} into an \textit{Advanced} group. Since the \textit{Novice} group represents individuals without formal training in cloud physics, comparing the \textit{Adept} and \textit{Advanced} groups investigates the key changes induced by increased academic experience in the discipline for learners. We created joint graphs by combining all individual graphs within a group (e.g., based on cloud physics experience). Nodes and edges represent the union of these elements from the individual graphs, with edge weights reflecting their repetition amongst participants.

A panel of four experts in cloud physics from the same Swiss teaching and research institution was consulted through interviews conducted after the analyses. They were asked about 1) what key topics and concepts can be progressively introduced to learners with increasing experience (Q1); 2) what main divergences in understanding there are between novices and experts, and what common misconceptions they identified in cloud physics (Q2); and 3) which most complex concepts of cloud physics need time to mature for learners (Q3). We then presented our findings to each expert for comment. Their inputs, presented in the Appendix, complement the Discussion section by offering informed perspectives on the levels of conceptual understanding revealed in teaching and learning interactions. These expert insights also strengthen the study’s findings through data triangulation, thereby enhancing the validity of the results. This data was also analysed using thematic content analysis \cite{Clarke2017} in a similar procedure to the one described above.

\section{Results\label{sec:results}}

In this section, the main results are presented. Discussion and interpretation of the results are provided in the subsequent sections.

\begin{figure}
\includegraphics[scale=0.4]{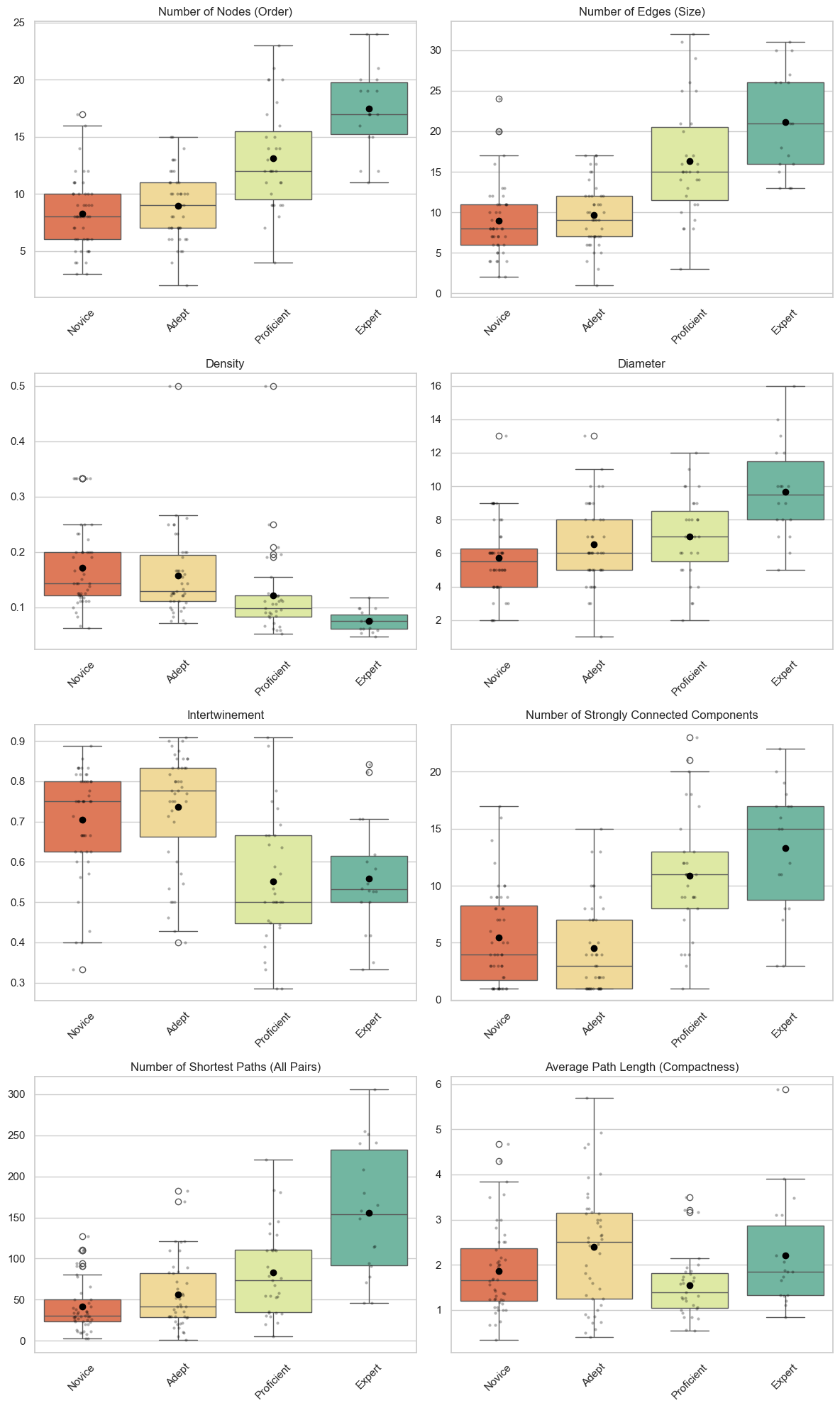}
\caption{Distribution of graph-level metrics for all participants across groups of cloud physics experience, inspired by Weihs and Drange \cite{Weihs2024ardist}. Boxes show the interquartile range, medians are marked by central lines, and black circles indicate the mean of each distribution.\label{fig:figure5}}
\end{figure}

Figure \ref{fig:figure5} presents the computation of several graph-level metrics for a grouping of participants by experience. The computations of the average degree, the number of cycles, and the average cycle lengths (not shown) indicated no significant differences between any groups. The boxes containing $50\%$ of the data points computed for these metrics overlapped across all experience groups.

\begin{figure}
\includegraphics[scale=0.36]{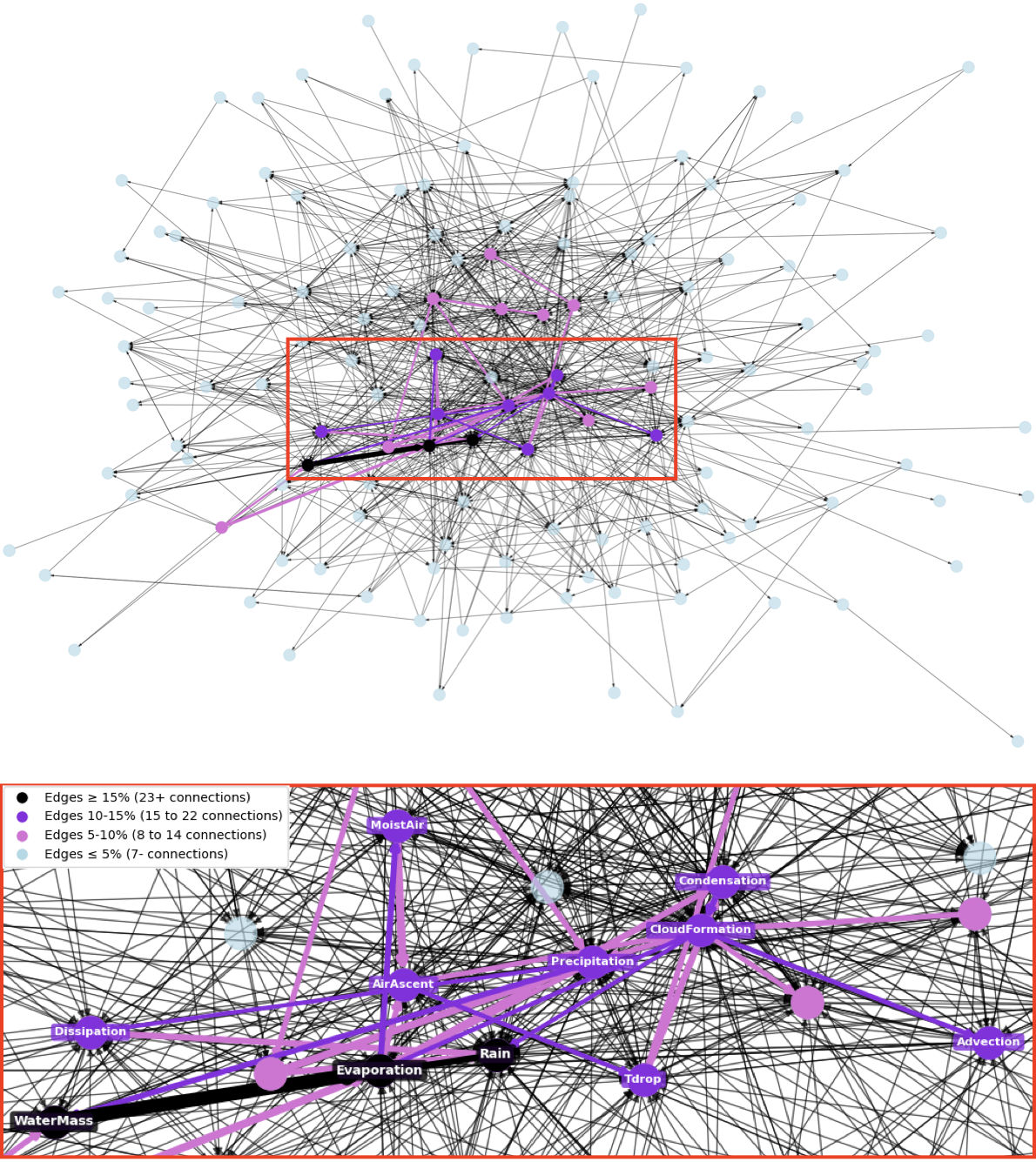}
\caption{Collective map of cloud physics understanding across all $153$ \textit{Novice}, \textit{Adept}, \textit{Proficient}, and \textit{Expert} participants, with agreement thresholds set at $5\%$, $10\%$ and $15\%$ of the participants regarding the edges (see legend for color coding). The direction of connections between the nodes are shown by arrows. The red rectangle on the top graph outlines the area of the bottom zoomed graph, and in which the nodes with highest agreement are labeled. \label{fig:figure6}}
\end{figure}

\begin{figure}
\includegraphics[scale=0.55]{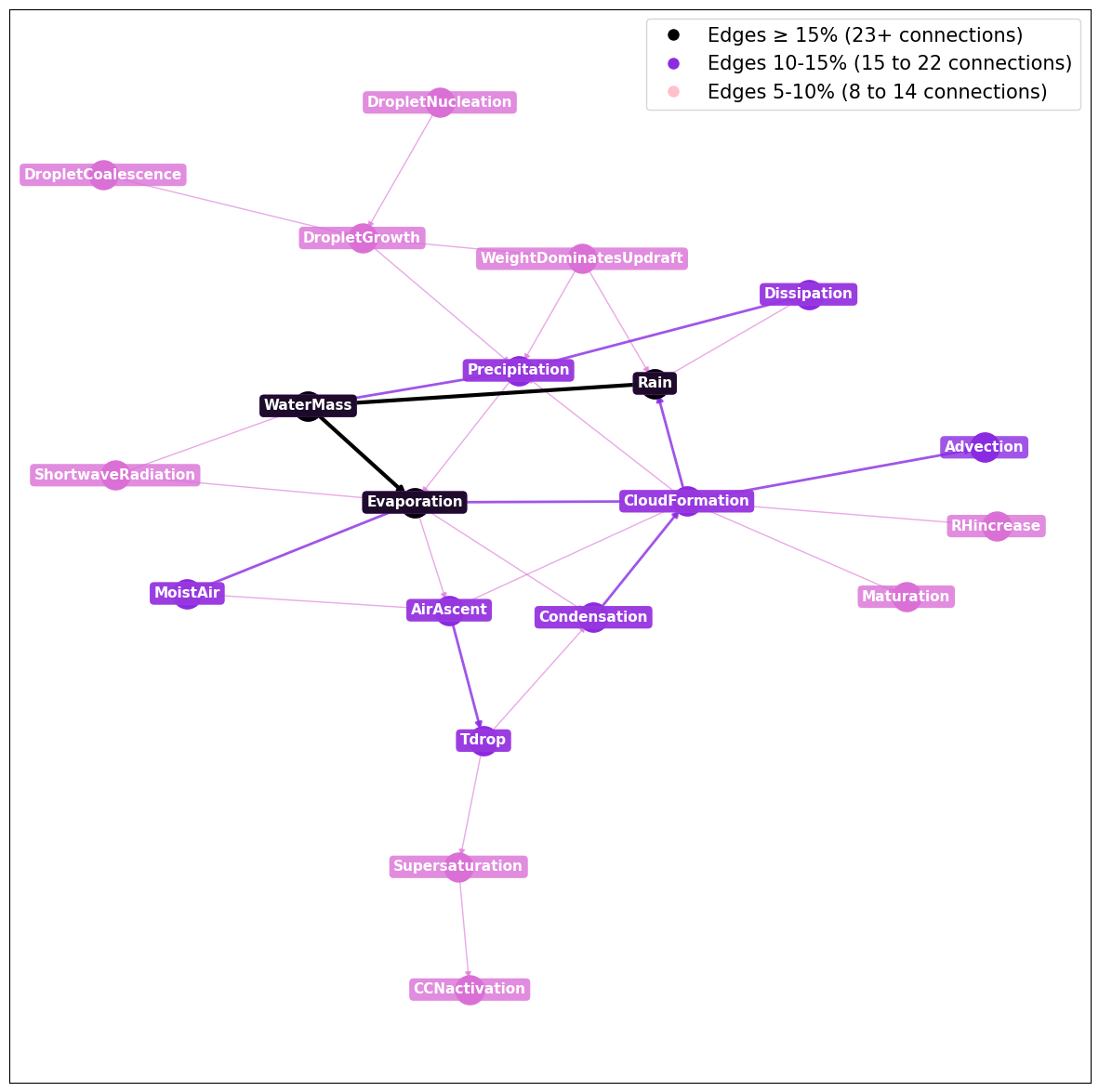}
\caption{Filtered and cleaned version of Figure \ref{fig:figure6}, only representing its colored subgraphs (see legend for color coding). \label{fig:figure7}}
\end{figure}

The joint weighted graph, shown in Figure \ref{fig:figure6}, presents the complete `map of cloud physics' derived from the data of $153$ participants, totaling $132$ concepts and $949$ weighted edges. A key objective of our study is to identify first-order structural patterns within this dense and complex network. By applying threshold values to edge weights, we isolate colored subgraphs that reflect varying levels of consensus across the entire sample. Representing only these subgraphs yields Figure \ref{fig:figure7}, in which the number of nodes is reduced to $20$. The most mentioned nodes across the map are \textit{Water Mass}, \textit{Evaporation} and \textit{Rain}. Other frequently used and highly connected nodes are \textit{Precipitation}, \textit{Dissipation}, \textit{Air Ascent}, \textit{Moist Air}, \textit{Temperature Drop}, \textit{Cloud Formation}, \textit{Condensation}, and \textit{Advection}.

\begin{figure}
\includegraphics[scale=0.36]{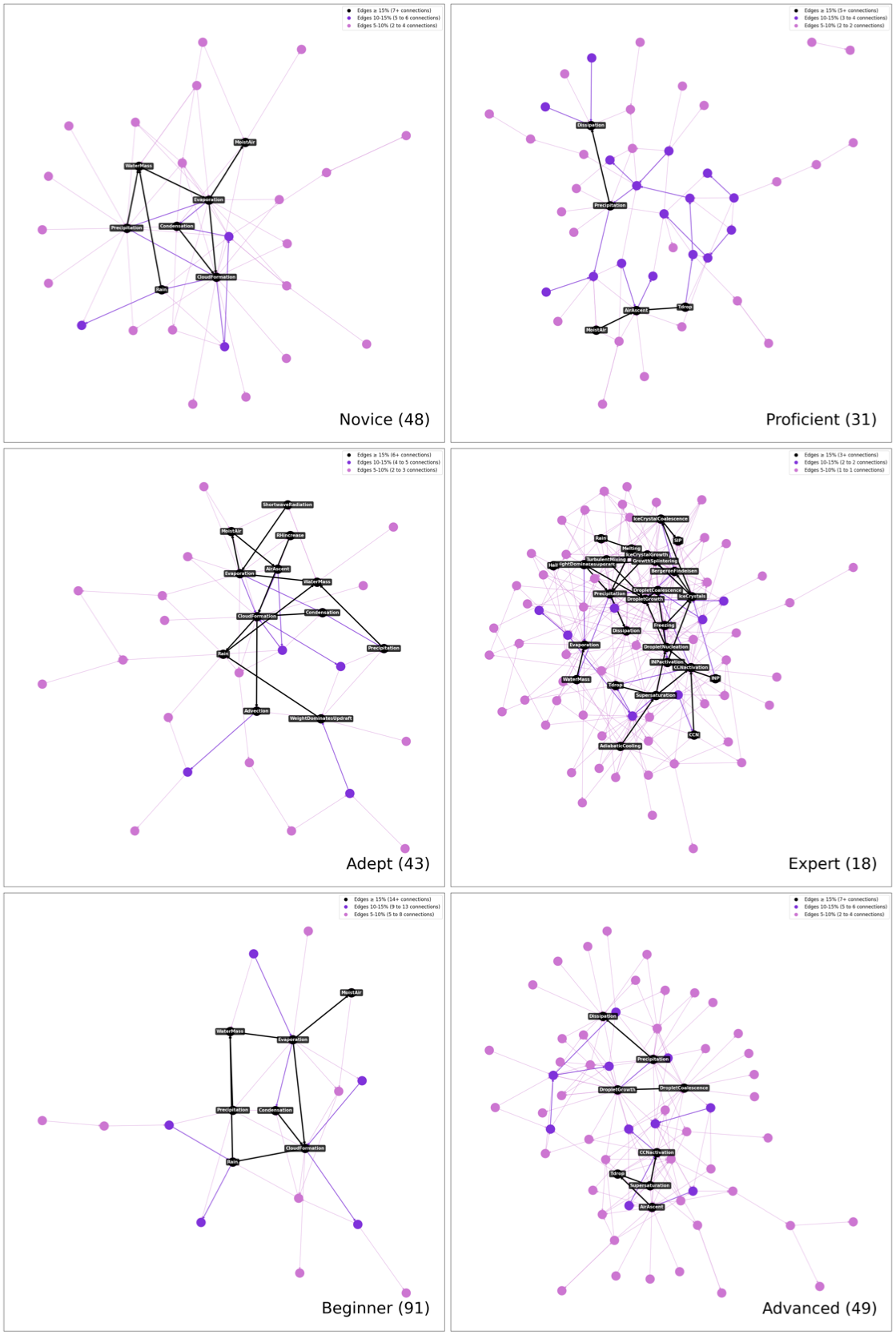}
\caption{As in Figure \ref{fig:figure7}, but for the \textit{Novice}, \textit{Adept}, \textit{Beginner} (\textit{Novice} \& \textit{Adept}), \textit{Proficient}, \textit{Expert}, and \textit{Advanced} (\textit{Proficient} \& \textit{Expert}) group subgraphs. \label{fig:figure8}}
\end{figure}

A similar procedure run on the graphs of all experience groups of participants leads to Figure \ref{fig:figure8}. The presented subgraphs represent the core knowledge of each group, with increasing levels of agreement shown in increasingly darker colors. All graphs of Figure \ref{fig:figure8} use the same threshold values of 5-10-15\%.

\begin{figure}
\includegraphics[scale=0.44]{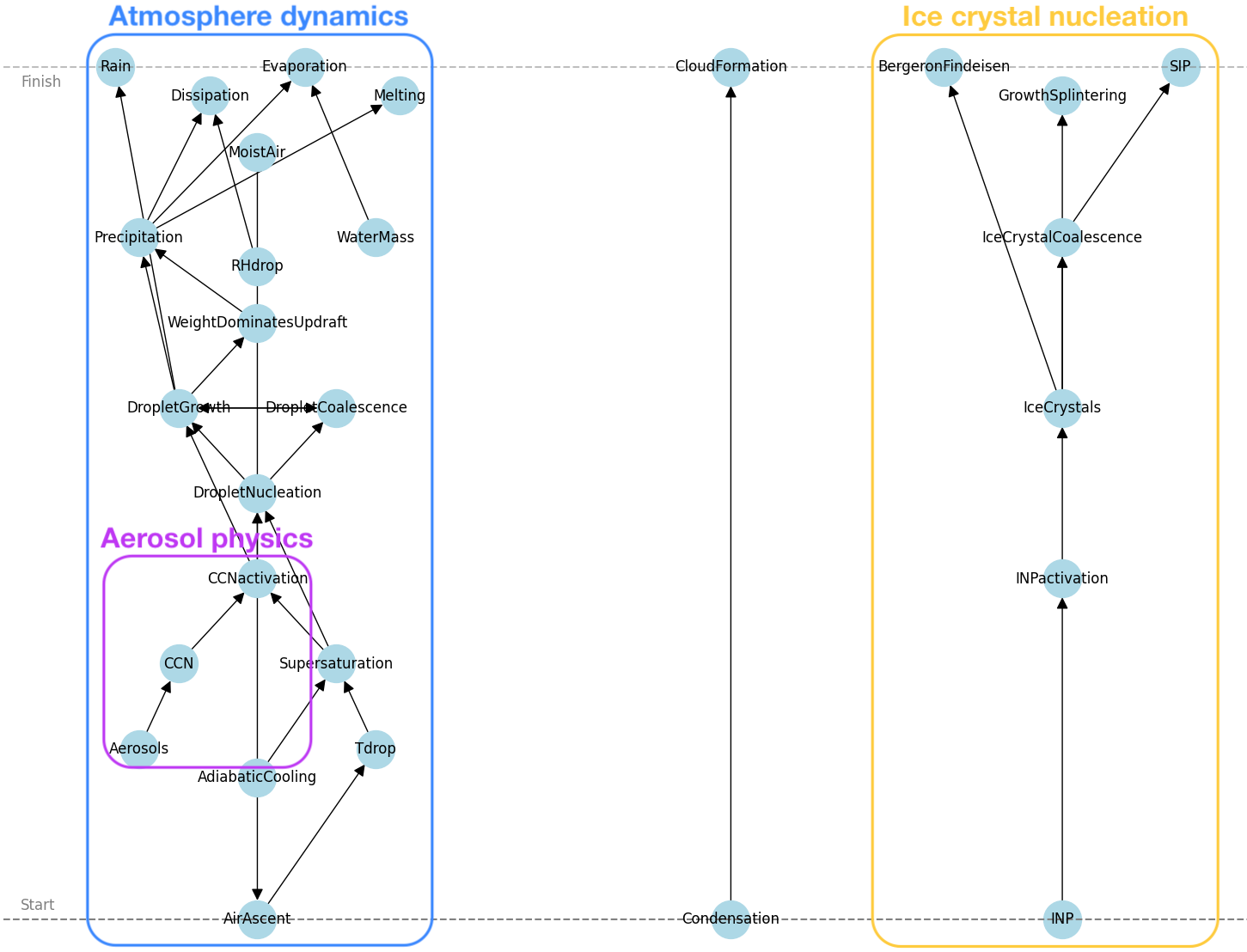}
\caption{Hierarchical core knowledge structure of the \textit{Advanced} group (49 participants). The three different structures are three subgraphs corresponding to subsets of the joint graph meeting at least 10\% agreement across the group, and represented by the violet and black structures in Figure \ref{fig:figure8}. The lines `Start' and `Finish' have been introduced for better readability, connecting nodes with in-degree zero to nodes with out-degree zero. Although \textit{Rain} is a subordinate concept \cite{Murphy1997} of \textit{Precipitation}, both were mentioned distinctively in the leftmost substructure. \textit{SIP} in the rightmost substructure is \textit{Secondary Ice Production}. The three colored areas are addressed in the Results section. \label{fig:figure9}}
\end{figure}

In Figure \ref{fig:figure9}, the \textit{Advanced} subgraph is filtered using a 10\% agreement threshold (corresponding to the violet and black edges from Figure \ref{fig:figure8}) and presents a structured way of navigating the graph, drawing on the hierarchical approach described by Tatsuoka \cite{Tatsuoka1986}. The resulting graph reveals a two-part substructure of cloud physics: 1) the evolution of warm clouds including atmospheric dynamics (generating supersaturation in moist air), aerosol physics (describing \textit{Cloud Condensation Nuclei} (\textit{CCN}) activation for droplet nucleation) and the mechanisms behind droplet growth and cloud dissipation (blue outlined frame); and 2) the processes specific to mixed-phase- and cold-cloud formation and maturation with \textit{Ice Nucleating Particles} (\textit{INP}) activation, ice crystal nucleation and growth (yellow frame). The sampled population also largely agreed on the connection from \textit{Condensation} to \textit{Cloud Formation}. If we separate the substructure 1) into the atmospheric dynamics (blue frame) and aerosol physics (purple frame) domains, we obtain a total of three parts, corresponding to the three separable areas of expertise in cloud sciences overall, reflected by the typical profiles of researchers in the discipline \cite{Kreidenweis2019}.

\begin{figure}
\includegraphics[scale=0.53]{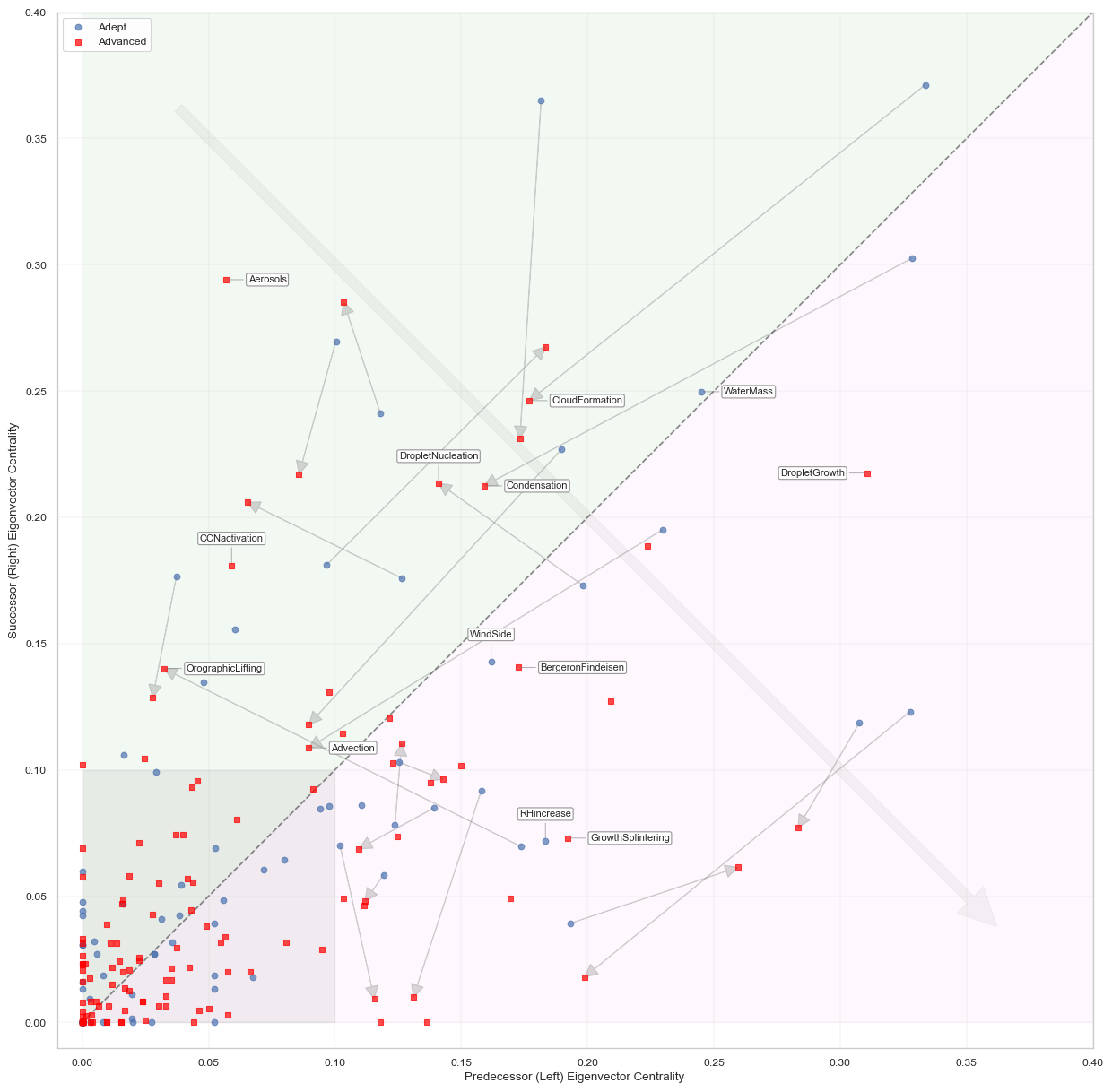}
\caption{Comparison of left and right eigenvector centralities from \textit{Adept} (blue) to \textit{Advanced} (red) groups. Only nodes where both the \textit{Adept} and the \textit{Advanced} values are outside the outlined box towards the origin are connected with \textit{Adept}-to-\textit{Advanced} arrows representing the gain in disciplinary experience. The gray diagonal timeline arrow is described in the Results section. \label{fig:figure10}}
\end{figure}

Node-level metrics were then computed for each group’s graph, in particular the left- and right-eigenvector, betweenness, and degree centralities. Figure \ref{fig:figure10} shows the evolution of the eigenvector centralities from the \textit{Adept} to the \textit{Advanced} groups, a measure that quantifies the importance of a concept at the beginning or at the end of an explanation.

For each node in the graph of either group, we calculated the left- (predecessor-based) and right-eigenvector (successor-based) centralities, which serve respectively as their $x$- and $y$-coordinates in Figure \ref{fig:figure10}. As these eigenvector centralities quantify the importance of nodes as relative starting or ending element in the participants’ description, the diagonal dashed line represents an area where left- and right-eigenvector scores are comparable. Nodes in this area are hence deemed important around the middle of a cloud’s life-cycle. These considerations allow us to draw a virtual timeline (in gray, with an arrow), where a progression from the green to the pink area represents a progression through the life-cycle of a cloud.

\begin{table}
\caption{Top 5 nodes scoring the largest difference from the \textit{Adept} to the \textit{Advanced} group, according to the degree and betweenness centrality metrics, with associated $z$-scores and $p$-values. The positive differences are displayed in bold, the negatives ones in italic.\label{tab:table1}}
\begin{ruledtabular}
\begin{tabular}{r l | r l}
\multicolumn{2}{c|}{Largest differences in degree centrality} &
\multicolumn{2}{c}{Largest differences in betweenness centrality} \\
\midrule
\textbf{Droplet Growth}        & \bm{$+0.31\ (3.31 \mid 0.001)$} & \textit{Cloud Formation}   & \textit{$-$0.19 (3.13 $|$ 0.002)} \\
\textit{Cloud Formation}      & \textit{$-$0.27 (3.21 $|$ 0.001)} & \textit{Water Mass}        & \textit{$-$0.13 (3.65 $|$ 0.001)} \\
\textit{Water Mass}            & \textit{$-$0.24 (5.94 $|$ 0.001)} & \textbf{Droplet Growth}    & \bm{$+0.10\ (2.02 \mid 0.038)$} \\
\textbf{Bergeron--Findeisen}  & \bm{$+0.23\ (3.78 \mid 0.001)$} & \textbf{Dissipation}       & \bm{$+0.07\ (1.38 \mid 0.199)$} \\
\textbf{CCN Activation}       & \bm{$+0.23\ (3.51 \mid 0.001)$} & \textit{Evaporation}       & \textit{$-$0.06 (1.19 $|$ 0.233)} \\
\end{tabular}
\end{ruledtabular}
\end{table}

Table \ref{tab:table1} presents the largest changes of the degree and betweenness centralities.

\begin{figure}
\includegraphics[scale=0.5]{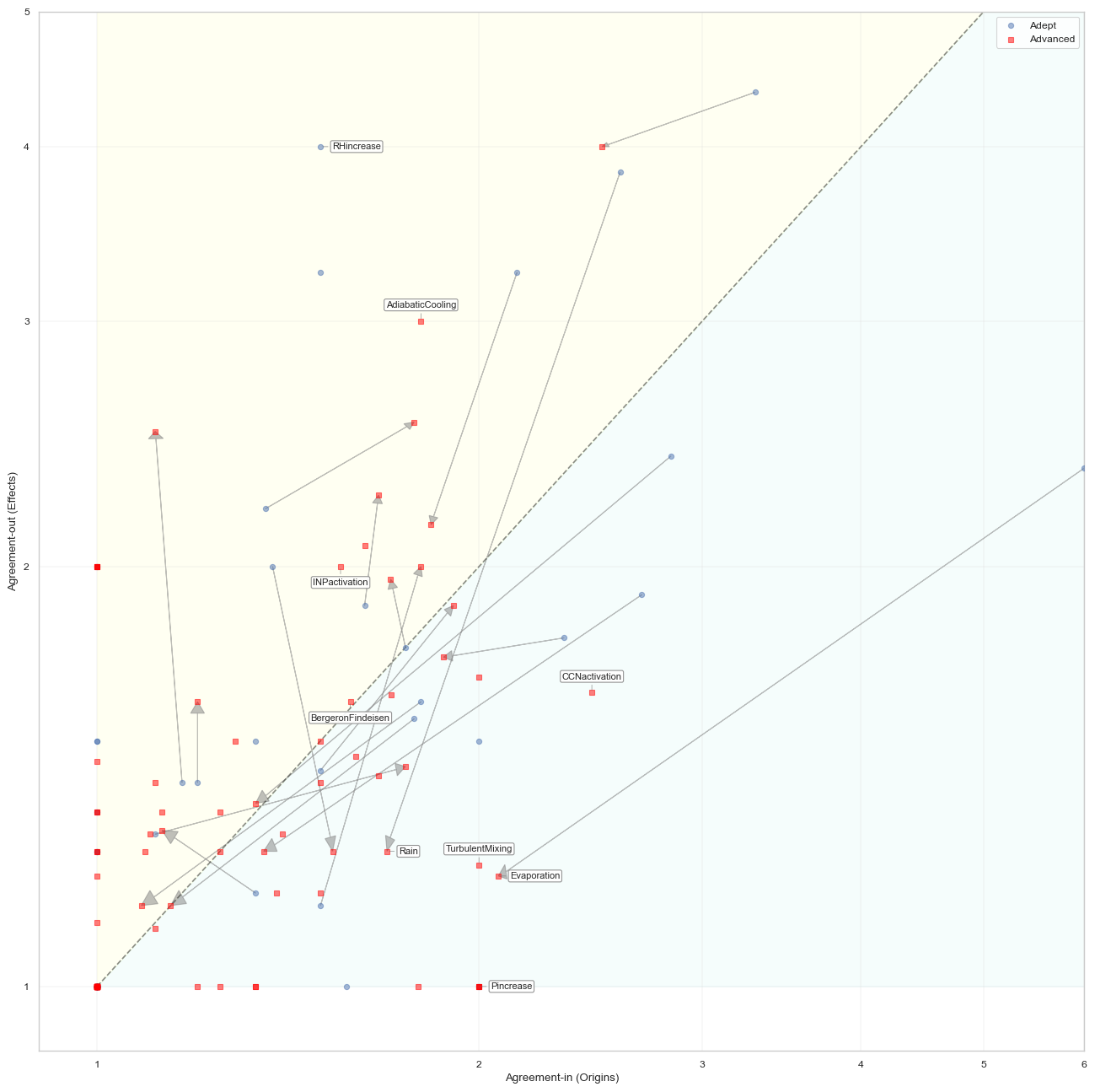}
\caption{Comparison of agreement-out and agreement-in metrics from \textit{Adept} (blue) to \textit{Advanced} (red) groups. Only nodes where both the \textit{Adept} and the \textit{Advanced} values are greater than $1$ on both axes are connected with \textit{Adept}-to-\textit{Advanced} arrows representing the gain in disciplinary experience. \label{fig:figure11}}
\end{figure}

The \textit{Adept}-to-\textit{Advanced} evolution of the agreement-in and agreement-out scores is presented in Figure \ref{fig:figure11}. This figure is created similarly to Figure \ref{fig:figure10}, with nodes in the yellow area being characterized by a higher agreement on their effects (the concepts connected out of them) than on their origins (the concepts connected into them), and vice versa for the turquoise area. Close to the dashed diagonal, the respondents agree on the origins as much as on the effects of the nodes.

\begin{table}
\caption{Top 5 nodes scoring the largest difference from the \textit{Adept} to the \textit{Advanced} group, according to the agreement-in and agreement-out scores, with associated $z$-scores and $p$-values. The positive differences are displayed in bold, the negative ones in italic.\label{tab:table2}}
\begin{ruledtabular}
\begin{tabular}{r l | r l}
\multicolumn{2}{c|}{Largest differences in agreement-in score} &
\multicolumn{2}{c}{Largest differences in agreement-out score} \\
\midrule
\textit{Evaporation}	& \textit{$-$3.18 (4.02 $|$ 0.001)} &	\textit{RH Increase}	& \textit{$-$3.00 (1.92 $|$ 0.055)} \\
\textbf{CCN Activation} &	\bm{$+2.45\ (4.72 \mid 0.001)$} &	\textit{Rain} &	\textit{$-$2.04 (1.92 $|$ 0.069)} \\
\textbf{Kinetic Energy Increase} &	\bm{$+2.00\ (1.47 \mid 0.491)$} &	\textbf{Cloud Thinning} &	\bm{$+2.00\ (1.40 \mid 0.489)$} \\
\textit{P Increase} &	\textit{$-$2.00 (1.73 $|$ 0.247)} &	\textbf{INP Activation} &	\bm{$+2.00\ (1.94 \mid 0.056)$} \\
\textbf{Droplet Breakup} &	\bm{$+2.00\ (1.53 \mid 0.481)$} &	\textbf{Adiabatic Cooling} &	\bm{$+2.00\ (2.13 \mid 0.023)$} \\
\end{tabular}
\end{ruledtabular}
\end{table}

Table \ref{tab:table2} shows similar \textit{Adept}-to-\textit{Advanced} results as Table \ref{tab:table1}, but for the agreement-in and agreement-out scores.

The Appendix presents the main themes emerging from the expert panel in response to the three questions asked, and the comments to the presented analysis results.
Furthermore, the Supplementary File 4, available to the reader upon request, displays heatmaps of the evolution of the centrality metrics and the timeline score.

\section{Discussion\label{sec:discussion}}

The results of the computation of graph metrics for participants grouped by cloud physics experience, presented in Figure \ref{fig:figure5}, are in line with the boxplots of Thurn et al. \cite{Thurn2020} and Wagner et al. \cite{Wagner2020}. Notably, while scholars identified expert-networks to be more ‘entangled’ \cite{Koponen2008} or ‘interconnected’ \cite{Siew2020} — which is synonymous with ‘intertwined’ \cite{Wagner2020} — their density is expected to be lower than that of novice-networks \cite{Thurn2020}. According to our analysis, disciplinary experience is the most statistically significant independent variable by which to group our participants. This contrasts with their academic discipline or level, strongly implying that exposure to a discipline is the key factor influencing the level of understanding of it, and confirming once more that deeper understanding depends on extended domain-specific knowledge, not the general education level or greater general abilities \cite{Chi2006}.

\subsection{Representing the conceptual understanding of cloud physics\label{subsec:representingconceptualunderstanding}}

The ‘map of cloud physics’, presented in Figure \ref{fig:figure6} and further filtered in Figure \ref{fig:figure7}, captures the shared understanding within a diverse learner population. It highlights the prevailing beliefs about the structure of the discipline. This subsection investigates the substructures that compose the map, both in light of RQ1 and to inform the disciplinary teaching and learning practices.

The joint graphs of each group (Figure \ref{fig:figure8}) serve as a tool to visualize the cloud discipline according to different levels of experience, and provide an empirical illustration of our network-based theoretical framework introduced in subsection \ref{subsubsec:networkframework}.  The \textit{Novice}, \textit{Adept} and \textit{Beginner} networks (left panel) are organized around relatively few concepts and relations, with prominent cycles centered around the chain of concepts $\textit{Water Mass} \rightarrow \textit{Evaporation} \rightarrow \textit{Cloud Formation} \rightarrow \textit{Rain} \rightarrow \textit{Water Mass}$. This chain replicates the main elements of the water cycle, both as it is being instructed by teachers \cite{Lee2019, Marquez2006, Vo2015} and understood by learners \cite{BenzviAssarf2005}. The highest-agreement structure of \textit{Novice} learners (in violet with the labeled nodes in Figure \ref{fig:figure8}), interpreted within our framework as reflecting their shared conceptual understanding, treats \textit{Rain} and \textit{Precipitation} as two distinct concepts contributing to \textit{Water Mass}. In addition, they connect \textit{Evaporation} to \textit{Moist Air}, although without explanatory continuation, suggesting a gap in understanding how to link \textit{Moist Air} to further stages leading to \textit{Cloud Formation}. Lastly, they connect to \textit{Cloud Formation} from both \textit{Evaporation} and \textit{Condensation}, which hints at a misconception about the nature of a cloud, as identified again in a later analysis in subsection \ref{subsec:analysingcharacterisingknowledgestructures}.

Conversely, the \textit{Proficient} and \textit{Advanced} graphs (right panel) contain more concepts and relations, and feature several substructures of highest agreement, such as $\textit{Precipitation} \rightarrow \textit{Dissipation}\text{, and } \textit{Moist Air} \rightarrow \textit{Air Ascent} \rightarrow \textit{Temperature Drop}$, the latter describing the concept of adiabatic cooling. The \textit{Advanced} group exhibits two distinct strands of understanding: one describing the conditions leading to cloud formation, and another describing the microphysics of cloud maturation, as further discussed below. The \textit{Expert} group presents a continuum of edges of highest agreement, indicating a strong consensus on a larger part of their descriptions. Also, the \textit{Expert} network clearly has fewer nodes with single connections than any other learner group, indicating that most elements in their description are integrated into a densely and coherently connected knowledge structure, consistent with the greater conceptual interconnectedness associated with expertise in our theoretical framework.

The highest-agreement structures shared across the groups in the right panel of Figure \ref{fig:figure8} suggest a qualitative shift in disciplinary experience. Recognizing such emerging structure could inform instruction that emphasizes the coherence among interrelated processes, rather than treating topics as discrete elements of the syllabus. Moreover, the progressive enrichment and reorganization of conceptual networks with increasing experience provide a potential basis for a learning progression in cloud physics, which we present here. Educational materials and activities could follow this progression, supporting a transition from descriptive toward relational reasoning that mirrors the epistemic development observed across experience levels.

\begin{figure}
\includegraphics[scale=0.75]{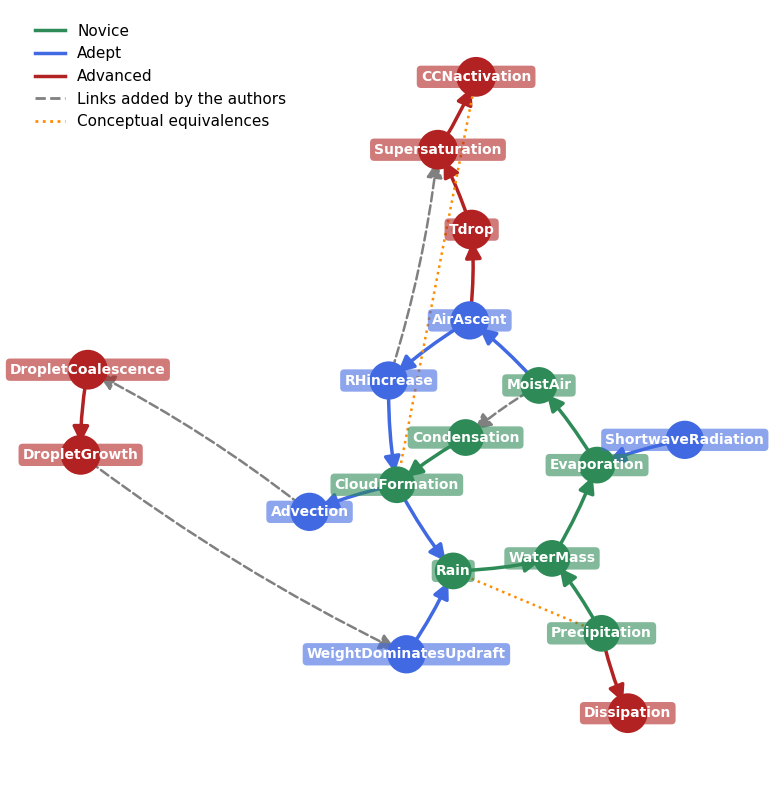}
\caption{Learning progression in cloud physics, tracing the substructures of highest agreement across \textit{Novice}, \textit{Adept} and \textit{Advanced} groups (see legend for color coding). The dashed directed edges in gray have been added by the authors to avoid disjoint elements. These connections are however validated by \textit{Expert} data. The two undirected edges in orange suggest conceptual synonyms, which could be merged without losing crucial information from the graph. \label{fig:figure12}}
\end{figure}

Synthesizing the progressive emergence of highest-agreement substructures across experience levels yields the following tentative learning progression in cloud physics:
\begin{itemize}[itemsep=0pt,parsep=0pt,topsep=0pt]
    \item first (\textit{Novice}): $\textit{Rain}\text{ and } \textit{Precipitation} \rightarrow \textit{Water Mass} \rightarrow \textit{Evaporation} \rightarrow \textit{Moist Air} \text{, and } \textit{Condensation} \rightarrow \textit{Cloud Formation}$;
    
    \item then (\textit{Adept}): $\textit{Moist Air} \rightarrow \textit{Air Ascent} \rightarrow \textit{RH Increase} \rightarrow \textit{Cloud Formation}\text{ and } \\ \textit{Weight Dominates the Updraft} \rightarrow \textit{Rain}\text{, as well as }\textit{Shortwave Radiation} \rightarrow \textit{Evaporation} \text{, and }\textit{Cloud Formation} \rightarrow \textit{Advection}$;
    
    \item finally (\textit{Advanced}): $\textit{Air Ascent} \rightarrow \textit{Temperature Drop} \rightarrow \textit{Supersaturation} \rightarrow \textit{CCN Activation} \text{, as well as } \textit{Precipitation} \rightarrow \textit{Dissipation}\text{, and } \\ \textit{Droplet Coalescence} \rightarrow \textit{Droplet Growth}$.
\end{itemize}
A graphical version of this progression is presented in Figure \ref{fig:figure12}.

The \textit{Advanced} core knowledge subgraph in Figure \ref{fig:figure9} presenting a hierarchy of concepts can serve as another recommendation to introduce the topic sequentially in a teaching and learning setting: cloud physics can be subdivided into three main sections: the dynamics of the atmosphere, the formation of liquid cloud droplets around aerosols, and the formation of ice nuclei around INP. Such a hierarchical graph can also be used to intertwine these different sections, for instance by setting the agreement threshold value accordingly: in our example of 10\%, atmospheric dynamics and aerosol physics are connected, but ice crystal nucleation is disjoint. These  observations are also in line with the expert panel suggestions regarding teaching progressions, recommending starting with thermodynamics and classical mechanics (blue in Figure \ref{fig:figure9}) before engaging with warm cloud formation (purple) and finally ice crystal
nucleation (yellow) (Appendix).
Based on data from learners with substantial coursework and disciplinary experts, the
Advanced graph from Figure \ref{fig:figure8} and Figure \ref{fig:figure9} can be considered as a reference structure to guide Novice and Adept learners, in line with evidence that expert-like network organization reflects higher-quality scientific understanding \cite{Wagner2023}.

\subsection{Analyzing and characterizing knowledge structures\label{subsec:analysingcharacterisingknowledgestructures}}

Our results provide both a qualitative and quantitative account of the epistemological shift in the description of a cloud’s life-cycle. Learners of lower levels of experience structure their explanations around the general physics of the water cycle, relying on concepts such as \textit{Water Mass}, \textit{Condensation}, \textit{Evaporation}, \textit{Rain} or \textit{Cloud Formation}. As they gain experience in the discipline, they increasingly use more detailed concepts that describe cloud microphysical processes, such as \textit{Droplet Growth}, \textit{Aerosols}, or \textit{Adiabatic Cooling}. Comparison of node-level metrics --- such as degree, betweenness and eigenvector centralities \cite{Koschutzki2008} --- across groups informs about the evolution of the roles that concepts have in the graphs of learners as they gain more experience in the discipline, and helps us answer RQ2. We present the concepts with most substantial metrics changes, and report their associated $z$-scores and $p$-values where these changes are largest.

Figure \ref{fig:figure10} shows how differences in eigenvector centrality between \textit{Adept} and \textit{Advanced} learners reveal a shift in the perceived importance of disciplinary concepts as experience increases. This is the case for \textit{Droplet Growth} gaining importance based on both its predecessors $(+0.24 \mid 3.33 \mid 0.001)$ and successors $(+0.16 \mid 2.31 \mid 0.016)$, reinforcing its core position in the life-cycle of a cloud. This also applies to microphysical concepts such as \textit{Aerosols}, \textit{CCN Activation} and the \textit{Bergeron--Findeisen [process]} as starting elements of an explanation, and for \textit{Growth by Splintering} as a finishing element. These findings are confirmed by the expert panel (Appendix), who note that \textit{Droplet Growth}, \textit{Aerosols} and \textit{CCN Activation} are active research topics and therefore receive considerable attention within expert groups. The panel also observed that less experienced learners generally overlook the importance of aerosols in cloud formation.

In contrast, the importance of general concepts decreases. \textit{Water Mass} shows a notable centrality decrease based on its predecessors $(-0.21 \mid 3.82 \mid 0.001)$ and successors $(-0.19 \mid 3.31 \mid 0.001)$, reducing its overall influence in the life-cycle of a cloud. Concepts such as \textit{Condensation} and \textit{RH Increase} lose influence as finishing elements of an explanation. The panel attributed the loss of influence of \textit{Water Mass} to the tendency of advanced learners to begin their descriptions with \textit{Evaporation}, implicitly assuming the presence of an initial water source.

We lastly compute the largest overall eigenvector centrality $C^{\text{overall}}_E$ changes by calculating the norm as described in the Supplementary File 1. The most significant changes occur for \textit{Water Mass} $(-0.29 \mid 3.56 \mid 0.001)$ and \textit{Droplet Growth} $(+0.29 \mid 2.91 \mid 0.003)$, \textit{Aerosols}, the \textit{Bergeron--Findeisen [process]} and the \textit{Wind Side [of a mountain]}. Changes for the finishing elements \textit{Advection} and \textit{Wind Side} could be influenced by local environments: many in the \textit{Adept} group live in mountainous regions (Norway), where orographic clouds—requiring wind and mountains for formation and precipitation—are a familiar phenomenon. By contrast, the \textit{Advanced} group has larger expertise with a variety of cloud types, many of which do not need to include these two concepts in their descriptions.

Interestingly, certain concepts also switch sides for their importance within the timeline of a cloud’s life-cycle; this is the case for \textit{Orographic Lifting}, \textit{Droplet Nucleation}, \textit{Advection} or \textit{Condensation}, that \textit{Adept} learners treat as finishing elements, whereas they appear earlier in \textit{Advanced} learners’ graphs. The influence of \textit{Orographic Lifting} and \textit{Advection} as finishing elements for \textit{Adept} learners supports the hypothesis of orographic clouds being the most potent representation for this group mostly located in Norway. The late \textit{Adept} placement of \textit{Condensation} hints at a possible misconception that \textit{Condensation} directly causes precipitation, whereas the \textit{Advanced} group places the concept early in the life-cycle, as a predecessor of \textit{Droplet Nucleation}. This suggestion, as well as the above hypothesis, could be the object of follow-up work focusing on misconceptions and geographically influenced representations of clouds.

The evolutions of degree and betweenness centralities (Table \ref{tab:table1}; Supplementary File 4) confirm the core role that \textit{Droplet Growth} gains in the \textit{Advanced} graph, both as a highly connected concept $(+0.31 \mid 3.31 \mid 0.001)$ and as an intermediary bridging other concepts $(+0.10 \mid 2.02 \mid 0.038)$. This dominant gain is supported by the expert panel, who identified
\textit{Droplet Growth} as the core intermediate step in a cloud’s life-cycle (Appendix).
Similarly, the \textit{Bergeron--Findeisen [process]} and \textit{CCN Activation} show increases in degree centrality. Although the \textit{Bergeron--Findeisen} process is at the interplay of many different processes, it is not used to describe a warm cloud, which is the cloud best understood by
\textit{Adept} learner. The panel also noted that \textit{Droplet Growth} and \textit{CCN Activation} feature in the description of hydrometeor growth, which replaces at higher experience level the more generic concept of \textit{Cloud Formation}. By contrast, \textit{Cloud Formation}, \textit{Water Mass} and \textit{Rain} show declines in connectivity and betweenness centrality, pointing at a change of vocabulary from these three central elements of the water cycle to more technical microphysical processes for the \textit{Advanced} group.

In their work, \cite{Koponen2014} argue that a concept property called ‘contingency’ refers to the different epistemic connections a concept has to others, and that
another one called ‘cohesion’ relates to cyclical patterns amongst concepts in a network. Both of these properties characterize what they propose as ‘key concepts’. As explained by the authors, key concepts would be of highest importance to a discipline, both from a teaching (because central to the field) and learning (because of their epistemic acceptability within a knowledge structure) point of view. In our study, the contingency is analogue to the degree centrality. We propose that betweenness centrality captures the other essential feature of a ‘key concept’; if the nodes of highest betweenness centrality were removed, the knowledge structure would be greatly affected in its coherence—other domains analyzing node-removal vulnerabilities according to betweenness centrality talk about ‘flow disruption’ or ‘critical damages’ to the network, see for example \cite{Demsar2007}. Therefore, we here use the degree and betweenness centrality scores as two important features of key concepts in cloud physics, and identify \textit{Droplet Growth}, \textit{CCN Activation}, and the \textit{Bergeron--Findeisen [process]} to be emergent key concepts in the discipline according to high experience learners.

The analysis of the agreement scores (Figure \ref{fig:figure11}; Table \ref{tab:table2}) shows that consensus on the origins and effects of most concepts decrease as learners gain experience. The most pronounced drops occur for the origins of \textit{Evaporation} $(-3.18 \mid 4.02 \mid 0.001)$ and \textit{P Increase}, and for the effects of \textit{RH Increase} $(-3.00 \mid 1.92 \mid 0.055)$ and \textit{Rain}. However, agreement increases with experience on the origins of \textit{CCN Activation} and the effects of \textit{INP Activation}, indicating that more experienced learners agree on the roles of these concepts in cloud processes.
When computing the overall evolution of agreement score, using a similar procedure as for the overall eigenvector centrality $C^{\text{overall}}_E$ (Supplementary File 1), we reveal the strongest significant decreases for \textit{Evaporation} $(-3.38 \mid 3.77 \mid 0.001)$ and \textit{RH Increase} $(-3.04 \mid 1.85 \mid 0.055)$. Meanwhile, agreement increases for \textit{CCN Activation}, \textit{Adiabatic Cooling} and \textit{INP Activation} (not shown), indicating a growing consensus on their functions amongst \textit{Advanced} learners. According to the expert panel (Appendix), these increases likely reflect recent scientific advances --- especially regarding INPs --- and the fundamental role of \textit{CCN Activation}, though possibly too complex for the \textit{Adept} learners to represent consistently.

The observed decreases in agreement could stem from the broader conceptual repertoire of \textit{Advanced} learners, which gives them more ways to explain processes and leads to a greater variation. However, decreasing agreement may also signal conceptual complexity: for instance, the roles of \textit{Evaporation} and \textit{RH Increase} are harder to define at \textit{Advanced}-level reasoning, according to the expert panel. By contrast, a decrease for general concepts such as \textit{Rain} and \textit{Shortwave Radiation} (not shown) likely relates to the fact that the \textit{Advanced} group did not use them in their descriptions.

Changes in timeline scores (Supplementary File 4) indicate that a gain in experience is accompanied by a relative increase in agreement on the effects of \textit{Frontal Wedging} $(\times 2.00 \mid 1.52 \mid 0.275)$, \textit{RH Drop}, and \textit{CCN} (note: $\times2.00$ denotes multiplication by a factor 2), and on the origins of \textit{RH Increase} $(\times0.38 \mid 1.04 \mid 0.943)$, \textit{Sensible Heat} and \textit{Condensation}. Although none of these results meet our thresholds for statistical significance, they reveal interesting patterns: the two extrema of the timeline score changes concern relative humidity: with greater experience, learners appear to converge on the origins of \textit{RH Increase} --- linked to \textit{Adiabatic Cooling} or \textit{Evaporation} --- and on the effects of \textit{RH Drop}, such as \textit{Dissipation} or \textit{Subsaturation}. This could also mean that the \textit{Advanced} group would agree less on the effects
of \textit{RH Increase}, for instance \textit{Air Ascent} or \textit{Small Droplets}, as revealed in the data, and on the origins of \textit{RH Drop}, for instance \textit{Downdraft} or different kinds of \textit{Precipitation}.

One limitation of the timeline score is its relative nature: increases may result from changes in the numerator, the denominator, or both, rendering the conclusions tentative. Even though timeline scores help highlighting emerging trends and patterns of conceptual understanding, we expect that expanding the sample size, particularly for the \textit{Advanced} group, would strengthen their explanatory power.

\section{Conclusion\label{sec:conclusion}}

In our study, we have shown that disciplinary experience is the criterion most influencing conceptual understanding in cloud physics. Addressing RQ1, we have presented graph-based visualization of a scientific discipline, revealing structures of highest agreement on its concepts and the relationships between them. In addition, we have identified substructures that are characteristic of the understanding of learners of different experience levels. To explore RQ2, we have demonstrated that the conceptual understanding in cloud physics develops from an initial focus around the water cycle (low experience level) to processes of cloud microphysics (high experience level). Also, we have identified the potential existence of regional biases in representations of clouds, as well as possible misconceptions amongst
low experience learners. Lastly, we flagged concepts suffering from low agreement amongst advanced learners, hinting at their potential inherent higher complexity. Taken together, these contributions advance both cloud physics education and the use of graph-based approaches in science education research.

Our findings are accompanied by recommendations formulated for teachers and learners of cloud physics. In particular, we derived a learning progression in the discipline, building up from novice-like chains of concepts to advanced-like conceptual structures. We also presented a teaching recommendation for the sequential introduction of the topics composing cloud physics. Finally, we highly encourage teachers to organize their instruction towards advanced-like knowledge structures, amongst others by emphasizing the role of \textit{CCN Activation} in cloud formation, the processes governing \textit{Droplet Growth}, and the \textit{Bergeron--Findeisen} ice crystal growth process.

\subsection{Future work\label{subsec:futurework}}

Regarding future work, the following directions are envisaged:
\begin{itemize}[itemsep=0pt,parsep=0pt,topsep=0pt]
    \item A Feature Reduction Analysis on all possible graph-level and node-level metrics would help identify the main variables explaining the largest variances across all the individual graphs’ metrics. This procedure could lead to the creation of a reduced set of variables used to run a clustering algorithm. These clusters could help validate the grouping by experience. First steps in this direction have been recently made by Weihs et al. \cite{Weihs2025paper3}.
    \item The use of the \textit{Advanced} data as a standard of conceptual understanding in the discipline could be a tool to assess the \textit{Novice} and \textit{Adept} data, and support the identification of common misconceptions at lower experience levels, building on work by Giabbanelli and Tawfik \cite{Giabbanelli2020}. Preliminary analysis made us detect that the ideas that ‘a cloud is gaseous’ or that ‘rain occurs on the leeward side of a mountain’ are misconceptions, which is in line with prior findings of atmospheric physics educational research \cite{Rappaport2009}.
    \item For educational purposes, supervised machine-learning classification of individual data based on our global analysis is an avenue to be explored. This could characterize a learner’s experience level, and highlight concepts where improved understanding would yield more expert-like structures, supporting personalized assessment and study recommendations for learners.
\end{itemize}

\subsection{Limitations\label{subsec:limitations}}

The findings of this study are subject to limitations. One aspect is the sampling mostly at institutions in one country, which may influence the results by eventual national scholarly traditions in the topic addressed. An expansion of the data collection to more countries could broaden the representations and characteristics of learners in the topic.

Another aspect is the short assessment time during the data collection. While we did not strictly enforce a time limit, most samples were collected within 15 min, which for some participants must have felt short to recollect and convert their knowledge into a concept map; either because the knowledge would be distant for some, or because the larger experience of others would have allowed them to contribute with an even richer input, given more time.

Similarly, the concept mapping exercise is likely unfamiliar to some. It is therefore expected that participants could have produced richer data given more experience with this kind of exercise.
The format of drawn maps with scientific concepts as nodes has been used previously by
scholars studying understanding. However, it remains limited compared to the depth
contained in natural language or other forms of data. A potentially great advancement to this study would be to design a way to convert interview data about a discipline into graph networks, and to analyze these much richer explanations. Similarly, the self-assessment of experience by the participants might not directly reflect their competencies.

We additionally acknowledge that the betweenness and degree centralities do not account for the weight edges. A method for integrating weights in their computation has been proposed by \cite{Singh2020} and could be the object of a future implementation.

Lastly, the absence of absolute consensus on the nature of the edges in this study (rather inferred subjectively from the collected data) leaves space for a scientific discussion on their interpretation.

\subsection{Implications\label{subsec:implications}}

In addition to the discussed implications for cloud physics education, the developed
methodology has the potential for broader application across STEM disciplines, for instance by collecting concept maps from learners on a subject in any STEM field, prompting them to ‘describe’ a larger process, and drawing on graph network analysis to process them as described in this study. The graph-based visualizations of a field may serve as valuable resources for both teachers and learners: both as a way to help teaching practitioners planning and structuring their lectures, and for students to use before and after attending their lectures, to retrieve previous knowledge, prevent the development of misconceptions, or support consolidating their knowledge structures. The employed methodology enables underlying structures to emerge from large and complex graphs, and reveals how an increase in disciplinary experience transforms the way learners interpret and navigate them.

\section{Acknowledgments\label{sec:acknowledgments}}

This research is supported by the Norwegian Center for Integrated Earth Science Education iEarth (\textit{Norwegian Agency for International Cooperation and Quality Enhancement in Higher Education} grant \#101060).
The authors would like to thank all the participants of this study for their contributions to the concept maps, and in particular Thomas Spengler, Hakan Heggernes and Martino Marisaldi at UiB, Carly Faber at UiT, Anders Mattias Lundmark at UiO, Rafael Kenji Horota, Knut Vilhelm Høyland, Lisa Baddeley, Tomasz Maciej Ciesielski and Frank Nielsen at UNIS, and Ulrike Lohmann at ETH Zurich for their support in the data collection. We also thank Ulrike Lohmann and her group for their precious insights into the conceptual understanding of cloud physics for learners of various levels, as well as Adrien Weihs at UCLA for helpful exchanges about graph theory algorithms. Lastly, we thank Christian Thurn at ETH Zurich and Steffen Wagner at the Humboldt University of Berlin for their careful rereading and valuable feedback on this manuscript.

\section{Declaration of interest statements and ethics statement\label{sec:declarationandstatements}}

The authors report there are no competing interests to declare. The study was approved by the Norwegian Agency for Shared Services in Education and Research SIKT (project number \#483904), and all the participants provided informed consent.

\section{Data availability statement\label{sec:dataavailability}}
The authors confirm that the data supporting the findings of this study are available in
Supplementary File 3, available to the reader upon request.

\newpage

\section{Appendix\label{sec:appendix}}

The sections below present the main ideas and emerging codes from the consultation with the expert panel about cloud physics concepts and our analysis results.

\subsection*{Q1 -- topic progression}

The experts identified the following examples of how understanding of cloud physics typically progresses with increasing expertise (represented by an arrow below).

{\footnotesize
\begin{itemize}[label=\textendash, itemsep=0pt, parsep=0pt, topsep=0pt, partopsep=0pt, leftmargin=1.5em]
    \item Droplet/crystal fall speed $\rightarrow$ Reynolds number, turbulence, and hydrodynamics.
    \item Hexagonal ice crystals $\rightarrow$ more complicated crystal shapes, dendrites, and vapor diffusion close to the crystal surface.
    \item Solid aerosols for ice/liquid droplets $\rightarrow$ cloud microphysics, ice nucleation processes, and thermodynamics of the atmosphere.
    \item Thermodynamics $\rightarrow$ Köhler theory.
    \item Ice/liquid cloud particles $\rightarrow$ droplet and ice-crystal formation and growth processes.
\end{itemize}}

\subsection*{Q2 -- Common differences in understanding}

The experts identified the following concepts as commonly representing differences between less experienced and more experienced learners.

{\footnotesize
\begin{itemize}[label=\textendash, itemsep=0pt, parsep=0pt, topsep=0pt, partopsep=0pt, leftmargin=1.5em]
    \item Köhler theory, including the superposition of the Raoult and Kelvin effects.
    \item Interpretation of Skew-T--$\ln p$ diagrams, including dry adiabatic ascent and constant moisture.
    \item Adiabatic cooling.
    \item Interpretation of observational cloud-droplet concentration data.
    \item The importance of supersaturation.
\end{itemize}}

\subsection*{Q3 -- Highest conceptual complexity}

The experts regarded the following concepts as among the most conceptually demanding ones in cloud physics.

{\footnotesize
\begin{itemize}[label=\textendash, itemsep=0pt, parsep=0pt, topsep=0pt, partopsep=0pt, leftmargin=1.5em]
    \item Ice-crystal nucleation.
    \item Köhler theory.
    \item Entrainment.
    \item Convective aggregation of updrafts and downdrafts.
    \item Stable boundary-layer clouds.
    \item Cooling resulting from evaporation, sublimation, and melting.
    \item Cloud-feedback mechanisms.
    \item Cloud dynamics.
    \item Ice-crystal growth and the development of columnar and hexagonal crystal habits.
\end{itemize}}

\subsection*{Expert feedback on the analysis results}

\newcommand{\analysisgap}{\vspace{0.4em}}

\begin{adjustwidth}{-2em}{0pt}

\textit{With more experience\ldots}

\textit{\ldots droplet growth gains a central role in the description of the cloud life cycle.}
{\footnotesize
\begin{itemize}[label=\textendash, itemsep=0pt, parsep=0pt, topsep=0pt, partopsep=0pt, leftmargin=4.4em]
    \item Droplet growth is currently an active topic of research and scientific uncertainty, receiving considerable attention from experts.
\end{itemize}}

\analysisgap

\textit{\ldots aerosols, CCN activation, and the Bergeron--Findeisen process become more important starting points.}
{\footnotesize
\begin{itemize}[label=\textendash, itemsep=0pt, parsep=0pt, topsep=0pt, partopsep=0pt, leftmargin=4.4em]
    \item Non-experts often do not realize that aerosols are required for cloud-droplet formation.
    \item Aerosols, CCN activation, and ice nucleation remain difficult to measure and therefore continue to be active areas of research.
\end{itemize}}

\analysisgap

\textit{\ldots water mass loses its central role in the description of the cloud life cycle.}
{\footnotesize
\begin{itemize}[label=\textendash, itemsep=0pt, parsep=0pt, topsep=0pt, partopsep=0pt, leftmargin=4.4em]
    \item Experts typically begin their explanations with evaporation rather than tracing the process back to water mass, which explains the reduced importance of this concept in expert networks.
\end{itemize}}

\analysisgap

\textit{\ldots condensation and advection become less important starting points.}
{\footnotesize
\begin{itemize}[label=\textendash, itemsep=0pt, parsep=0pt, topsep=0pt, partopsep=0pt, leftmargin=4.4em]
    \item Different cloud types may explain this observation. Orographic clouds require mountains and therefore make advection more salient for less experienced participants.
    \item Most Adept participants were from Norway, where orographic clouds are common, which may explain the stronger emphasis on advection.
\end{itemize}}

\analysisgap

\textit{\ldots droplet growth, the Bergeron--Findeisen process, and droplet coalescence become more strongly connected to other concepts.}
{\footnotesize
\begin{itemize}[label=\textendash, itemsep=0pt, parsep=0pt, topsep=0pt, partopsep=0pt, leftmargin=4.4em]
    \item Droplet growth represents the central intermediate stage of the cloud life cycle and therefore credibly connects to many other concepts and processes.
    \item Bergeron--Findeisen, droplet coalescence, and CCN activation all describe hydrometeor growth, replacing the more generalist 'cloud formation' as the central concept.
    \item The Bergeron--Findeisen process lies at the intersection of many cloud-physics processes, although it is not required to describe warm clouds.
\end{itemize}}

\analysisgap

\textit{\ldots droplet growth and convection become essential milestones.}
{\footnotesize
\begin{itemize}[label=\textendash, itemsep=0pt, parsep=0pt, topsep=0pt, partopsep=0pt, leftmargin=4.4em]
    \item The slower emergence of convection could depend on the cloud types considered by adepts.
    \item Nevertheless, convection remains an essential component to describe droplet nucleation.
\end{itemize}}

\analysisgap

\textit{\ldots cloud formation, water mass, and freezing lose their roles as essential milestones.}
{\footnotesize
\begin{itemize}[label=\textendash, itemsep=0pt, parsep=0pt, topsep=0pt, partopsep=0pt, leftmargin=4.4em]
    \item The reduction in betweenness centrality for cloud formation probably reflects changes in terminology rather than a genuine decrease in conceptual importance.
\end{itemize}}

\analysisgap

\textit{\ldots agreement increases for adiabatic cooling (origins) and INP activation (effects).}
{\footnotesize
\begin{itemize}[label=\textendash, itemsep=0pt, parsep=0pt, topsep=0pt, partopsep=0pt, leftmargin=4.4em]
    \item Adiabatic cooling represents fundamental atmospheric thermodynamics, whereas recent advances in research have increased consensus regarding the role of ice-nucleating particles.
\end{itemize}}

\analysisgap

\textit{\ldots agreement decreases for evaporation and pressure increase (origins), and increases for relative-humidity, rain, shortwave radiation, and evaporation (effects).}
{\footnotesize
\begin{itemize}[label=\textendash, itemsep=0pt, parsep=0pt, topsep=0pt, partopsep=0pt, leftmargin=4.4em]
    \item Adept participants may primarily associate increasing relative humidity with cloud formation, whereas experts additionally connect it to concepts such as CCN activation and Köhler theory.
    \item Experts possess many more possible conceptual connections and therefore differ more in the level of detail used to describe cloud life cycles.
    \item Evaporation, cloud formation, moist air, and shortwave radiation remain topics with important unresolved scientific questions.
\end{itemize}}

\end{adjustwidth}

\newpage

\bibliography{CloudPhysicsLibrary}

\end{document}